\documentstyle[11pt,newpasp,twoside,epsf]{article}
\begin{document}

\title{Stellar Nuclear Rings in Barred Galaxies: Fossils of Past 
Circumnuclear Starbursts?}
\author{Peter Erwin, Juan Carlos Vega Beltr\'an, John Beckman}
\affil{Instituto de Astrof\'{\i}sica de Canarias\\
38200 La Laguna, Tenerife, Spain}

\begin{abstract}
We have found four barred S0 galaxies --- NGC 936, NGC 3945, NGC 4340,
and NGC 4371 --- which contain smooth, luminous, purely stellar
nuclear rings within their bars.  These rings have little or no dust,
no evidence for recent star formation, and are approximately the same
color as surrounding bar and bulge.  Thus, they are probably the aged
remnants of bar-driven circumnuclear starburst episodes similar to
those seen in barred galaxies today.

Using kinematic data from long-slit spectroscopy, we construct
rotation and resonance curves for two of the galaxies.  In both cases,
the nuclear rings appear to be located near or at the inner inner
Lindblad resonances of the large-scale bars.

We also discuss the difficulties inherent in detecting and identifying
such rings, and show some of the surprising ways in which stellar
rings can distort galaxy isophotes and ellipse fits.
\end{abstract}

\section{Introduction}

Nuclear rings are circular or elliptical rings of gas and dust,
typically $\sim 1$ kpc in size, found inside the bars of nearby
galaxies; they are believed to result from the interplay of bar-driven
gas inflow and bar resonances (e.g., Athanassoula 1992; Piner, Stone,
\& Teuben 1995).  They are often sites of vigorous star formation. 
However, the fate of star-forming nuclear rings --- what happens after
the gas is consumed and star formation ceases --- is currently
unclear, since to date almost no ``fossil'' nuclear rings are known 
(the extremely small, blue ring found by van den Bosch \& Emsellem 
1998 in NGC 4570 is a possible exception).

We have recently found evidence for four such rings, in the SB0
galaxies NGC 936, NGC 3945, NGC 4340, and NGC 4371.  The sizes of
these rings --- 600 to 900 pc in radius and 9--12\% the size of the
respective bars --- are consistent with those of typical ``young''
nuclear rings (Buta \& Crocker 1993).  These rings are detected only
as smooth distortions of the stellar isophotes; there is no evidence
for significant dust or recent star formation.

\section{Finding Fossil Nuclear Rings}

Young (i.e., gas-rich and star-forming) nuclear rings are rather easy
to see, particularly in color maps (e.g., Buta \& Crocker 1993).  But
an old ring can be much harder to detect, since it may not show up in
color maps and may be hard to discern against a bright bulge. 
Figure~1 shows a simplistic model galaxy, consisting of a de
Vaucouleurs bulge of ellipticity 0.25 and a Gaussian ring of
ellipticity 0.6.  The ring distorts the isophotes in two ways: it
makes the isophotes more elliptical at its radius, and the isophotes
are rounder and ``boxy'' just inside the ring.  Looking at just the
isophotes, or at isophotal ellipse fits, it would not be clear what
causes such a distortion; it could be mistaken for, e.g., a nuclear
bar.  However, the presence and shape of the ring is easily recovered
through unsharp masking (bottom half of figure).  The similarity to 
the real galaxy NGC 4371 is striking.

\begin{figure}
\plotone{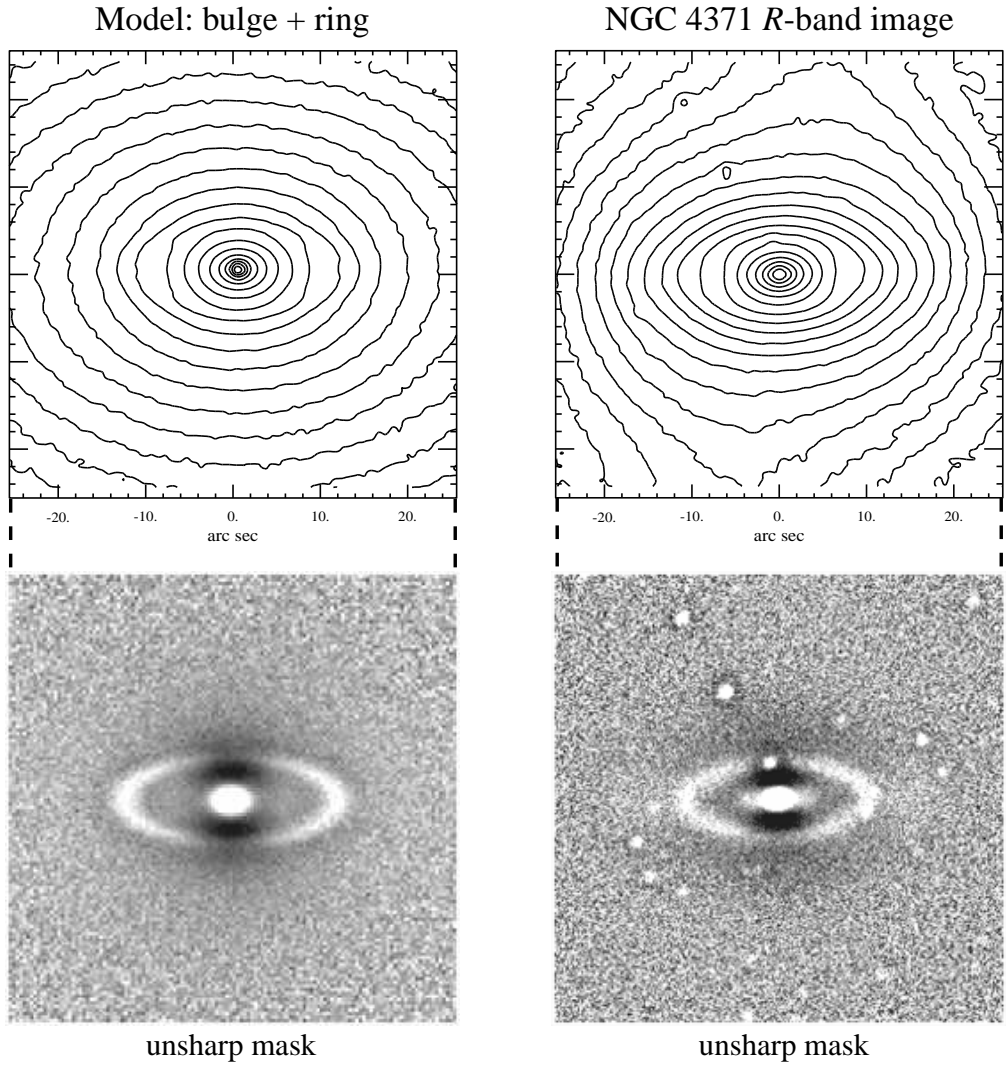}

\caption{Left: model image consisting of a moderately elliptical
$R^{1/4}$ bulge and a more elliptical ring, with isophotes above and
unsharp mask below.  Right: central $4 \times 4$ kpc of SB0 galaxy NGC
4371, with $R$-band isophotes above and unsharp mask below.}

\end{figure}

\section{Morphology and Color of Nuclear Rings}

In three of the galaxies (NGC~936, NGC~4340, and NGC~4371), the rings
are smooth, elliptical, and aligned with the outer galactic disk: they
appear to be intrinsically circular.  In NGC~3945, the ring seems
intrinsically elliptical, and shows signs of spiral (possibly $m = 4$)
structure.  In both NGC~3945 and NGC~4340, small secondary bars, not
aligned with the large-scale primary bars, are found inside the
nuclear rings.  This pairing of nuclear rings and inner bars is also
seen in galaxies with young nuclear rings (e.g., Wozniak et al.\
1995).

Color maps show that rings are essentially ``colorless'': they are
indistinguishable from the surrounding bulge and bar stellar
population, and are thus probably old and dust-free.  The ring in
NGC~4371 may be an exception, since it is (slightly) bluer than its
surroundings; however, this could also be caused by a faint dust ring
outside the stellar ring (cf.\ Wozniak et al.\ 1995).

\section{Bar Resonances}

Modeling of gas flow in barred galaxies (Athanassoula 1992; Piner,
Stone, \& Teuben 1995) shows that gaseous nuclear rings are closely
associated with the inner Lindblad resonance(s) (ILRs) of large-scale
bars.  We can define approximate locations for such resonances in our
galaxies by using kinematic information from long-slit spectroscopy;
this lets us see whether the stellar nuclear rings are also associated
with ILRs.

For NGC~4340, we use the major-axis velocities of Simien \& Prugniel
(1997) to derive a frequency curve (Figure~2, top).  We use our
(unpublished) bar-axis velocities, which extend to larger radii, to
estimate the outer bar¹s rotation speed: we assume that corotation
is at 1.1--1.3 times the bar length, as appears to be the case for
those SB0 galaxies with known bar pattern speeds (e.g., Merrifield \&
Kuijken 1995; Gerssen, Kuijken, \& Merrifield 1999; Elmegreen et al.\
1996).  The points where the bar pattern speed is equal to $\Omega -
\kappa/2$ are possible locations of the bar¹s ILR(s).  We do the same
for NGC~3945 (Figure~2, bottom), using Kormendy¹s (1982) major-axis
rotation curve, although our estimate of the bar pattern speed is more
uncertain due to the fact that the rotation curve does not extend past
the outer bar radius.

For both NGC~3945 and NGC~4340, it appears that the nuclear rings lie
near or at the inner ILR of the primary bars.  This strengthens our
identification of these rings as the stellar remnants of gaseous,
star-forming nuclear rings.

The resonance curves also indicate that the inner bars in both of
these double-barred galaxies end at or inside the inner ILRs.  This is
consistent with theories of double-barred galaxies where corotation of
the faster-rotating inner bar coincides with one of the outer bar¹s
ILRs (Pfenniger \& Norman 1990; Maciejewski \& Sparke 2000).

\begin{figure}
\plotone{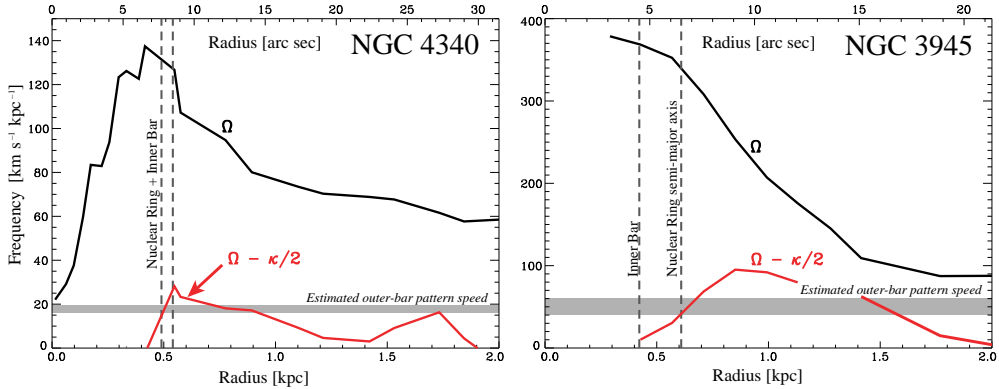}

\caption{Resonance curves for NGC 4340 and NGC 3945.  Vertical dashed
lines mark the deprojected sizes of inner bars and nuclear rings in
each galaxy; the horizontal gray bands show the estimated ranges of
outer-bar pattern speeds.  Possible locations of inner Lindblad
resonances (ILRs) are where the estimated bar pattern speed = $\Omega
- \kappa/2$ (where the $\Omega - \kappa/2$ curves cross the
pattern-speed bands).}

\end{figure}

\end{document}